\documentclass[conference]{IEEEtran}
\IEEEoverridecommandlockouts
\usepackage{cite}
\usepackage{amsmath,amssymb,amsfonts}
\usepackage{algorithmic}
\usepackage{graphicx}
\usepackage{textcomp}
\usepackage{xcolor}

\usepackage{float}
\usepackage{arydshln}
\usepackage{pifont} 
\usepackage{url}

\usepackage{epsfig}

\usepackage{multirow}
\usepackage{booktabs}
\usepackage{threeparttable}
\usepackage{subcaption}
\usepackage[hidelinks]{hyperref}

\def\BibTeX{{\rm B\kern-.05em{\sc i\kern-.025em b}\kern-.08em
    T\kern-.1667em\lower.7ex\hbox{E}\kern-.125emX}}

\begin{document}
\bstctlcite{IEEEexample:BSTcontrol}

\title{CrossMuSim: A Cross-Modal Framework for Music Similarity Retrieval with LLM-Powered Text Description Sourcing and Mining}

\author{
    \IEEEauthorblockN{Tristan Tsoi$^{1*}$\thanks{$^{*}$Equal contribution.}, Jiajun Deng$^{1*}$, Yaolong Ju$^1$, Benno Weck$^2$, Holger Kirchhoff$^3$, Simon Lui$^1$}
    \IEEEauthorblockA{$^1$Audio Lab Hong Kong, Hong Kong Research Center, Huawei Technologies, Hong Kong SAR}
    \IEEEauthorblockA{$^2$Music Technology Group, Universitat Pompeu Fabra, Barcelona, Spain}
    \IEEEauthorblockA{$^3$Munich Research Center, Huawei Technologies, Munich, Germany}\{tsoi.tik.nang1, deng.jiajun1, yaolongju, holger.kirchhoff, luisiuhang\}@huawei.com
}

\maketitle

\begin{abstract}
Music similarity retrieval is fundamental for managing and exploring relevant content from large collections in streaming platforms. This paper presents a novel cross-modal contrastive learning framework that leverages the open-ended nature of text descriptions to guide music similarity modeling, addressing the limitations of traditional uni-modal approaches in capturing complex musical relationships. To overcome the scarcity of high-quality text-music paired data, this paper introduces a dual-source data acquisition approach combining online scraping and LLM-based prompting, where carefully designed prompts leverage LLMs' comprehensive music knowledge to generate contextually rich descriptions. Extensive experiments demonstrate that the proposed framework achieves significant performance improvements over existing benchmarks through objective metrics, subjective evaluations, and real-world A/B testing on the Huawei Music streaming platform.
\end{abstract}


\vspace{-0.1cm}
\section{Introduction}
\label{sec:intro}
Music similarity retrieval plays an important role in many music information retrieval (MIR) tasks, such as music recommendation \cite{siamese}, personalized playlist generation \cite{inferring} and background music replacement in video editing \cite{Disentangled, MusicalAudioSimilarity}. As digital music collections rapidly expand within streaming platforms, accurately identifying similarities between musical pieces has become critical for managing and exploring relevant content from such large collections efficiently. For example, when a user expresses preference for a particular song, the systems operate on a query-by-example basis where one song serves as a reference to retrieve similar music content.  

Previous work on music similarity retrieval has primarily spearheaded into two distinct areas: 1) Annotation-based retrieval strategies leverage diverse descriptors, including textual editorial metadata (e.g., artist \cite{artist} or titles \cite{semantic_annotation}), manual annotations (e.g., genre \cite{mpeg7} or tempo \cite{social_tagging}), automated music tagging \cite{autotagging, collecting_tags} (e.g., automatic genre recognition or tempo estimation) or visual elements (e.g., album artwork \cite{retrieval_album_covers} or band photographs \cite{judge_an_artist}), to represent the actual music content and perform retrieval by analyzing and comparing such information. 2) In audio content representation-based retrieval approach, audio content embeddings are extracted directly from music recordings via either digital signal processing \cite{improvements, function, AudioContent} or deep learning techniques \cite{MusicalAudioSimilarity, Perceptual, Disentangled, siamese, deep_ranking}, where these embeddings serve as the basis for retrieving similar musical content. For example, contrastive learning is utilized to learn an embedding for each music audio recording, bringing similar music audio closer together and pushing dissimilar audio further apart in the latent representation space \cite{chen2020simple}.
These latent representation-based approaches, specifically those utilizing contrastive learning framework \cite{Perceptual, Disentangled}, not only overcome the scalability limitations and potential biases inherent in annotation-based approaches, but also benefit from their ability to capture acoustic similarity directly from audio content. 

Existing contrastive learning approaches have explored various learning targets for similarity modeling, such as defining acoustic similarity between augmented versions of the same audio \cite{MusicalAudioSimilarity, Perceptual}, or between audio samples sharing predefined categorical attributes (e.g., genre, tempo, or instrument) \cite{Disentangled}. However, these uni-modal contrastive approaches remain limited, as their reliance on basic augmentation schemes and limited expressiveness of predefined, closed-form categorization fundamentally constrains their ability to capture the complex, nuanced relationships inherent in musical content. The widespread success of cross-modal text-music contrastive learning in a range of downstream tasks, such as music autotagging, text-to-music retrieval \cite{toward, mulan, muscall, WikiMuTe, mulap}, and text-to-music generation \cite{musiclm, stable} suggests its potential for music similarity modeling, where natural language descriptions offer richer semantic information than traditional predefined tags.

However, the acquisition of high-quality semantic-relevant text descriptions presents substantial challenges for musical modeling in practical applications. Although music clips are generally accessible, their associated textual descriptions are often scarce or difficult to obtain. Furthermore, even when text data can be obtained from music videos or online sources, a critical challenge lies in identifying and selecting qualified textual content that meaningfully correlates with the underlying musical concepts we aim to model, such as genre, mood, tempo and other structural elements.

\begin{figure*}[h!]
    \centering
    \includegraphics[width=\textwidth]{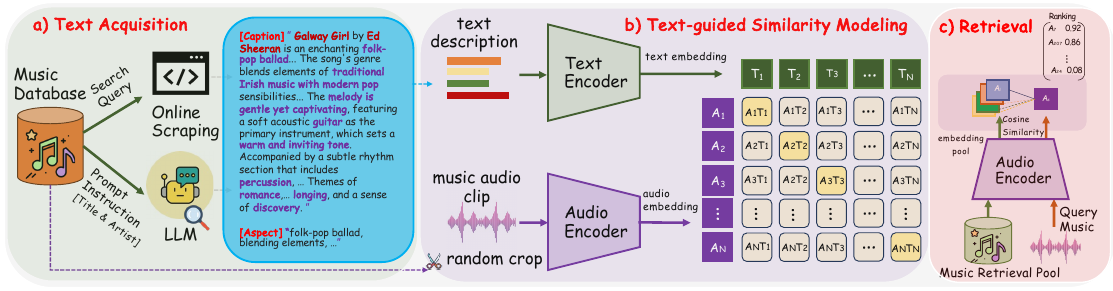} 
    \caption{Overview of the proposed CrossMuSim framework. (a) Dual-source data acquisition module combining online scraping and LLM-based prompting for textual description generation, (b) Music similarity modeling utilizing cross-modal contrastive learning framework with text-music pairs, and (c) Inference phase for music similarity retrieval using audio modality.}
    \vspace{-0.2cm}
    \label{fig:system_diagram}
    \vspace{-0.4cm}
\end{figure*}

To this end, this paper presents a novel \textbf{cross}-modal framework for \textbf{mu}sic \textbf{sim}ilarity retrieval (CrossMuSim) that utilizes text-music contrastive learning. The approach leverages the expressive power of natural language descriptions associated with music tracks to model underlying musical concepts. To address the textual data scarcity, we introduce a complementary data acquisition strategy to flexibly source a broad range of music using accessible resources: \textbf{a)} an online data mining technique searching beyond typical English sources, and \textbf{b)} a prompting methodology that employs large language models (LLMs) to generate diverse, informative descriptions using basic music identifiers. The contributions of this paper are summarized as follows:

\textbf{1)} This paper presents a cross-modal contrastive learning framework that leverages the open-ended nature of free-form text to guide music similarity modeling. Existing approaches to music similarity modeling mainly focus on manual annotations \cite{artist,semantic_annotation,mpeg7} and uni-modal representation learning with pre-defined labels \cite{siamese, deep_ranking}, restricting their scalability and expressiveness. While recent cross-modal contrastive learning has shown promise in unified embedding learning \cite{clap, mulan} and text-to-music retrieval \cite{mulan,WikiMuTe,toward,muscall}, its potential for music-to-music retrieval task remains unexplored.

\textbf{2)} This paper presents a dual-source data acquisition approach combining online scraping and LLM-based prompting to address text-music data scarcity. Unlike previous works that used LLMs merely for paraphrasing \cite{lpmusiccaps}, summarization \cite{clamp2}, and caption generation with extensive keywords \cite{augment, lpmusiccaps, laion}, our carefully designed prompts leverage LLMs' extensive knowledge of well-known songs and artists to generate contextually diverse descriptions for better similarity modeling using only song identifiers as cues, similar to \cite{jonason_2023_10178563}.

\textbf{3)} 
The effectiveness of the proposed cross-modal framework for music similarity modeling is demonstrated over extensive evaluations, including both objective and subjective evaluations, as well as online A/B testing on the Huawei Music platform for downstream music recommendation.

\section{Text-Music System}
\label{sec:system}
\subsection{Architecture}
\label{sec:architecture}
Figure \ref{fig:system_diagram}(b) illustrates the architecture of our text-music system \cite{WikiMuTe}, comprising a text encoder and an audio encoder, each with a projection layer that aligns their outputs to the same dimension. The projection layer is a lightweight component consisting of a two-layer MLP and rectified linear units (ReLU) activation applied between the layers. For the \textbf{text encoder}, a sentence transformer model \cite{sentence} optimized for generating sentence-level embeddings is utilized to model long dependencies of text. More specifically, we opt for the distiluse-base-multilingual-cased-v2 model, which is aligned across languages to effectively process multilingual textual data from Chinese and English sources. For the \textbf{audio encoder}, the compact Music Tagging Transformer (MTT)  model \cite{mtt} which is pre-trained on the Million Song Dataset \cite{million} for the music auto-tagging task is adopted. 

The representation of the text encoder is obtained by applying mean pooling to the output sequence, while for the audio encoder we use the embedding of the special [CLS] token from the encoder’s output.

\subsection{Contrastive Learning Framework}

 A sample consists of a (text, audio) pair. We minimize a variant of the InfoNCE loss, namely the Normalized Temperature-scaled Cross Entropy (NT-Xent) loss \cite{loss}, to align their representations in a unified latent space. For a batch containing $N$ samples, the loss $\mathcal{L}$ is expressed as

\begin{equation}
\label{eq:loss}
\mathcal{L}_{\text{NT-Xent}} = \sum_{i=1}^{N} \log \frac{\exp(z_{i,i}/\tau)}{\sum_{j=1}^{N} \exp(z_{i,j}/\tau)}
\end{equation}

\noindent where $z_{i,j}$ represents the cosine similarity between the text of the i-th sample and the audio of the j-th sample and $\tau$ is a temperature hyperparameter.

\section{Data Sourcing and Mining}
\label{sec:mining}
In downstream MIR tasks such as music recommendation, user preferences are often diverse and span a wide range of musical genres, styles, and contexts, it is essential to explore scalable and flexible data sourcing approaches for capturing such variability. We explore two methods for acquiring high-quality text-music data: (i) a semi-automated online scraping pipeline to collect text data, and (ii) leveraging LLMs to generate informative descriptions.

\subsection{Online Scraping}
\label{sec:scraping}

In the context of online scraping, we source music descriptions from a Chinese web encyclopedia (CWE) with dedicated music pages that encompass a wide variety of music genres. The search queries consist of a comprehensive set of terms spanning genre (e.g., ``pop"), mood (e.g., ``romantic"), theme (e.g., ``party"), language (e.g. ``Japanese"), timbre (e.g., ``full-bodied"), and tempo (e.g., ``slow"). For each piece of music retrieved, we scrape its metadata, background, and song review sections. The song review section provides detailed descriptions that often integrate multiple aspects of the music, along with additional context such as historical background and MIR characteristics. These entries are typically curated by experienced contributors, ensuring high-quality content.

\subsection{LLM-based Prompting}
\label{sec:generation}

High-quality text data with diverse, rich descriptions of musical elements is inherently limited in quantity. In contrast, metadata, such as song titles, artist names, or album names, is abundant and readily accessible online. Developing methods to generate content-rich textual data from minimal inputs can further facilitate modeling on more diverse music datasets. Inspired by \cite{lpmusiccaps, augment}, we investigate the retrieval of music-related information from LLMs by querying them using only music identifiers limited to artist name and song title.

Specifically, we explore the feasibility of this approach using two state-of-the-art (SOTA) compact LLMs, GPT-4o-mini and Qwen2-7B. To minimize LLM hallucination and thus ensure output reliability, we curate an audio dataset of 27k \textit{well-known} songs sourced exclusively from popular playlists (e.g. ``Billboard Year-End Hot 100 singles", ``Mandopop Top 100") and prominent artists (e.g., Adele, Taylor Swift, Jay Chou), limited to works released up to 2020. This dataset is subsequently used to prompt the LLMs to generate musical descriptions using only the song title and artist name. The prompt is specifically designed to elicit semantic information about key musical characteristics, including genre, melody, instrumentation, tempo, vocal characteristics, mood, thematic elements, and lyrics\footnote{Data, prompt, and demo examples are available at: \url{https://crossmusim.github.io/}}. An example of an LLM-generated output is shown in Figure~\ref{fig:system_diagram}(a).

\subsection{Aspect Generation}
Keywords are extracted from the scraped/generated descriptions using LLMs to create aspects that provide a concise, high-level summary of each musical piece. These compact textual representations not only align with the annotation style of open-source datasets \cite{musiclm, WikiMuTe} but also potentially facilitate more effective learning in the latent space.

\section{Experiments}
\subsection{Datasets}
\label{sec:datasets}
Apart from the web-sourced data collections, we also utilize open-source and in-house datasets for training and evaluation. Two music semantics-focused open datasets are utilized, \textbf{a) MusicCaps} \cite{musiclm} consists of 5.5k 10-second music clips from the AudioSet dataset. Each clip is annotated with short descriptors and captions written by musicians. \textbf{b) WikiMuTe} \cite{WikiMuTe} contains 9k music tracks paired with music descriptions.

In addition, we utilize an in-house collection of 200k songs selected among the platform's most popular tracks,  which have all been manually labeled by professional musicians. The tagging system is organized in a multi-level hierarchy, ranging from coarse-grained labels to fine-grained labels (e.g., ``genre" $\rightarrow$ ``pop" $\rightarrow$ ``sentimental pop"). For this study, we focus on three relevant musical aspects: genre, mood, and scenario\footnote{Scenario refers to the context in which a song is deemed suitable for consumption, which may be related to activity, weather, season, time of day, location, or sport type.}, and select songs that contain at least one tag in each of these dimensions, resulting in a subset of 150k samples. To mitigate the data imbalance issue, tags with fewer than 500 occurrences were excluded from the dataset.

\begin{table*}[h!]
\caption{Performance of the baseline models, online scraping methods, and the enhanced approaches combining online scraping with LLM-based prompting using various text formats in terms of text-to-music (T2M) and music-to-music (M2M) retrieval metrics. ``$^\ast$" denotes the open-source datasets. ``$\dagger$" represents evaluation using mAP@100. ``$\diamondsuit$" means closed-source.}
\centering
\resizebox{1.0\textwidth}{!}{ 
\begin{tabular}{@{}lllllccccc@{}}
\toprule
\hline
\multirow{2}{*}{ID} & \multirow{2}{*}{Method} & \multirow{2}{*}{Dataset} & \multirow{2}{*}{\multirow{2}{*}{\shortstack[c]{Text Description\\ Format} }} & \multirow{2}{*}{\multirow{2}{*}{\shortstack[c]{Captioning/LLM\\ Model} }} & \multicolumn{3}{c}{Text-to-Music}  & \multicolumn{2}{c}{Music-to-Music$^\dagger$}  \\ \cmidrule(l){6-10} 
  &          &   &  &  & R@1& R@5& mAP@10 &Coarse &  Fine \\ \cmidrule(r){1-5} \cmidrule(r){6-8} \cmidrule(r){9-10}
1 &  Random  & - & - & - &-   & -  &  -     &  25.98   &  14.28   \\
2 &  CLMR \cite{clmr}  & - & - & - &-   & -  &  -     &  42.90   &  35.91   \\
3 &  MERT-v0 \cite{mert}  & - & - & - &-   & -  &  -     &  39.13   &  30.07   \\
4 &  MTT \cite{mtt} & - & - & - &-   & -  &  -     &  \textbf{47.33}   &  \textbf{50.59}   \\
5 &  WiKiT2M$^{\diamondsuit}$ \cite{WikiMuTe}  & - & - & - &-   & -  &  -     &  46.99   &  45.03   \\ \midrule
6 &  \multirow{3}{*}{\shortstack[l]{Online\\Scraping} }  &  \multirow{3}{*}{CWE}   & Caption & - & 0.08   & 0.59  &  1.07     &  46.38   &  50.99   \\ 
7 &    &   & Aspect & - & 0.14   & 0.77  & 1.28     &  \textbf{47.32}   &  49.37   \\ 
8 &    &   & Caption \& Aspect & - & \textbf{0.14} & \textbf{0.82}   & \textbf{1.30}  &  46.53  &  \textbf{52.64}   \\ \midrule
9 &  \multirow{2}{*}{\shortstack[l]{ Online\\Scraping} }  &  $\text{CWE+WikiMuTe}^\ast$  & \multirow{2}{*}{Aspect} & - & 0.17   & 1.07  &  1.41     &  47.44   &  50.25   \\ 
10 &   &  + $\text{MusicCaps}^\ast$  &  & - & \textbf{0.64}   & 2.39  &  3.03     &  47.87   &  51.64   \\  \cdashline{2-10}[1pt/2pt] 
11 &  +Captioning & \multirow{3}{*}{\shortstack{ ++ 27k In-house Data} }  & \multirow{3}{*}{Aspect} & LP-MusicCaps & 0.34   & 1.88  &  2.50      &  47.26   &  48.87   \\ \cdashline{2-2}[1pt/2pt]
12 &   \multirow{2}{*}{\shortstack[l]{+LLM-based \\ Prompting} }  &   &  & Qwen2-7B & 0.53   & 2.30  & 2.99     &  48.20   &  54.90   \\ 
13 &    &   &  & GPT-4o-mini& 0.63   & \textbf{2.46}  & \textbf{3.21} &  \textbf{48.24}   &  \textbf{53.10}   \\ \midrule
14 &  \multirow{2}{*}{\shortstack[l]{Online\\Scraping} }  &  $\text{CWE+WikiMuTe}^\ast$  & \multirow{2}{*}{Caption \& Aspect} & - & 0.29   & 0.93  &  1.44     &  47.09   &  53.35   \\ 
15 &   &  + $\text{MusicCaps}^\ast$  &  & - & \textbf{0.64}   & 2.40  & 3.11     &  47.47   &  53.59   \\  \cdashline{2-10}[1pt/2pt] 
16 &  +Captioning & \multirow{3}{*}{\shortstack{ ++ 27k In-house Data} }  & \multirow{3}{*}{Caption \& Aspect} & LP-MusicCaps & 0.48   & 1.90  &  2.54     &  46.86   &  50.02   \\ \cdashline{2-2}[1pt/2pt]
17 &  \multirow{2}{*}{\shortstack[l]{+LLM-based \\ Prompting} }  &   &  & Qwen2-7B & 0.53   & 2.14  & 2.79    &  47.84   &  58.41   \\ 
18 &    &   &  & GPT-4o-mini& 0.63   &  \textbf{4.40} & \textbf{3.25} &  \textbf{47.99}   &  \textbf{58.72}   \\ \hline\bottomrule
\end{tabular}
}
\label{tab:big}
\vspace{-0.3cm}
\end{table*}

\subsection{Experimental Settings}

\textbf{Data Preprocessing:} The training process can include one or more types of text, such as short aspects and long captions, with each text type treated as a distinct training sample. To generate text from aspects, up to five aspects are chosen randomly from the aspect list. For captions, the text is split into sentences and a random block of sentences is selected from the start. Descriptions are sentence-tokenized, while lyrics are segmented by line breaks. Each text sample is paired with a 10-second audio segment randomly sampled from the recording.

\textbf{Experimental Setup}: The encoders are loaded with pre-trained weights while all projection layers are randomly initialized. The system is optimized using Adam, with the learning rate warmed up linearly to 1e-4 over one epoch, followed by a cosine decay scheduler applied over 40 epochs. The temperature $\tau$ for the loss function is set to 0.07. 
Training is conducted with a batch size of 64 across 4 GPUs. Inference is performed on the entire audio clip in non-overlapping 10-second intervals, with intermediate outputs mean-pooled to produce the final representation.

\vspace{-0.1cm}
\subsection{Baselines}
To the best of our knowledge, none of the existing music similarity works are open-source, which motivates our selection of SOTA music encoders from various domains as benchmarks: \textbf{a) CLMR} \cite{clmr}: A well-established benchmark built on a self-supervised contrastive learning framework, leveraging self-augmentation techniques and fine-tuned for music autotaggng; and \textbf{b) MERT} \cite{mert}: A versatile music foundation model that has demonstrated robust performance across 14 downstream MIR tasks. The MERT-v0 base model is adopted. In addition, we benchmark against the pre-trained \textbf{c) MTT} \cite{mtt} and \textbf{d) WiKiT2M} \cite{WikiMuTe} models, as well as a \textbf{e) random} retrieval baseline.

\vspace{-0.1cm}
\subsection{Evaluation Metrics}

\textbf{Text-to-Music Retrieval}: We perform tag-based queries \cite{WikiMuTe} on the evaluation set. Retrieval performance is measured using standard metrics, including recall at $k$ (R@$k$) and mean average precision at $k$ (mAP@$k$), where $k$ presents the top-$k$ ranked items. The evaluation split from MusicCaps \cite{musiclm} and the Chinese benchmark MuChin \cite{muchin} are combined to create a multilingual evaluation set. 

\textbf{Music-to-Music Retrieval}: Following the approach in \cite{MusicalAudioSimilarity}, we perform query-by-example retrieval and compute mAP@100 for all tags, where songs sharing the same tag are considered positives. To create the evaluation set, 50k samples are randomly selected from the in-house collection. Unlike previous works that focus solely on a single retrieval level \cite{improvements, Disentangled, MusicalAudioSimilarity}, we examine both 1) a coarse-grained level that classifies broader categories reflecting listening suitability, including genre, mood, and scenario and, 2) a fine-grained level that focuses on specific subcategories within genres, such as distinct pop types, chosen for the central role of genre in organizing music collections \cite{improvements}.

\textbf{Subjective Evaluation}: We conduct a Mean Opinion Score (MOS) evaluation for music-to-music retrieval. For each query, the system retrieves three similar songs from the evaluation set. 10 professional musicians then evaluate the quality of the retrieval based on three criteria: musical similarity, acoustic similarity, and overall similarity.

\textbf{Online A/B Testing}: A week-long A/B test is conducted on a personalized playlist feature on the Huawei Music platform, monitoring play counts and device coverage to gauge user's preferences. Traffic is split between two groups: A) using WiKiT2M \cite{WikiMuTe} and B) using the proposed method. During each user session, the system retrieves the 20 most similar songs from the 200k in-house collection for three ``liked" items, which are then ranked by a proprietary system to generate a session playlist.

\subsection{Results}
\textbf{Performance of the proposed CrossMuSim:} Table \ref{tab:big} shows the text-to-music and music-to-music retrieval results. Several trends can be observed. \textbf{a)} The consistent correlation between text-to-music and music-to-music performance validates that cross-modal contrastive learning facilitates music similarity modeling by learning audio-text associations in a unified latent space. \textbf{b)} Systems that use aspects alone achieve the best music-to-music retrieval performance at the coarse-grained level (ID.7 and ID.9-13), whereas combining aspects and captions yields the best performance at the fine-grained level (ID.8 and ID.14-18). \textbf{c)} Incorporating additional open datasets enhances performance and enables our system to outperform all the baselines (ID.9-10 and ID.14-15). This supports our assumption that high-quality paired data is fundamental to effective text-music modeling. \textbf{d)} Systems incorporating the LLM-based prompting technique outperform those relying solely on online scraped data (ID.12-13 and ID.17-18). Both GPT-4o-mini and Qwen2-7B demonstrate competitive improvements. Notably, the GPT-4o-mini variant trained on captions and aspects outperforms MTT by \textbf{0.66\%} and \textbf{8.13\%} absolute at the coarse-grained and fine-grained levels respectively, and are selected for the subsequent experiments. This improvement may be attributed to the robust outputs by the LLMs, which exhibit a consistent writing format and style, feature a diverse vocabulary, and align well with the musical content. These results highlight the feasibility of leveraging vast amounts of unlabeled raw audio data for dataset construction through an effective LLM-based prompting framework.

\begin{figure}[h!]
\vspace{-0.2cm}
    \centering
    \includegraphics[width=0.48\textwidth]{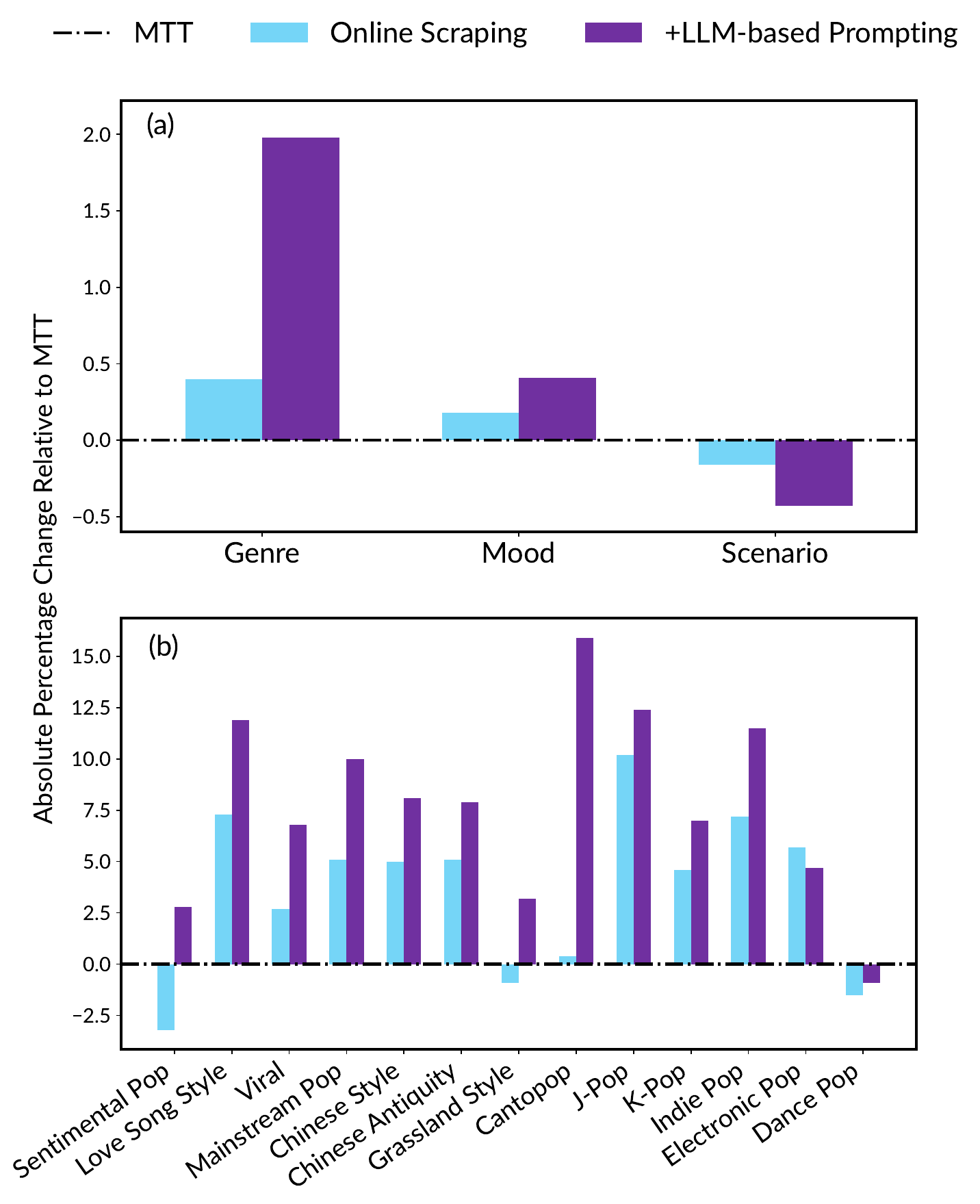} 
    \caption{Music-to-music performance of MTT baseline, online scraping, and online scraping with LLM-prompting methods in terms of a) coarse-grained and b) fine-grained music categorization levels.}
    \label{fig:comparison_chart}
\end{figure}

\textbf{Performance on the coarse-grained and fine-grained level categorizations:} Figure \ref{fig:comparison_chart} illustrates the absolute percentage improvements of the two proposed data sourcing methods over the MTT baseline. The following findings can be observed. \textbf{a)} Genre shows the largest improvement at the coarse-grained level, followed by mood, while scenario exhibits a negative impact. This can be attributed to the subjective and multifaceted nature of the mood and scenario categories. For example, a sentimental ballad might be labeled as ``sad" and evoke a ``rainy" atmosphere but could also be associated with other labels like ``deep feeling" or ``overcast.". This highlights a potential mismatch between the descriptors of the data sourcing methods and those emphasized in the evaluation. \textbf{b)} Online scraping with LLM-based prompting outperforms MTT in 12 out of 14 music types at the fine-grained level. In particular, significant improvements of \textbf{15.57\%}, \textbf{12.24\%}, and \textbf{7.54\%} are obtained over MTT on Cantopop, J-pop, and K-pop genres. A brief experiment is conducted using a music captioning model from \cite{lpmusiccaps}, to generate chunk-specific descriptions based on audio content at 10-second intervals, but it leads to suboptimal performance. We note that the captions often contain mismatched and inconsistent descriptions.

\textbf{Performance of subjective evaluation} are shown in Table \ref{tab:mos}, where online scraping and online scraping with LLM-based prompting significantly outperform MTT, achieving average MOS scores of \textbf{3.62} and \textbf{4.16}, respectively. These two methods are preferred across all dimensions, with particularly strong performance in musicality, as they are reported to effectively capture subtle genre and language differences across various styles of music. A similar trend is observed in the A/B testing results presented in Table \ref{tab:ab}. Specifically, playback count increases by \textbf{3.44\%} and \textbf{6.20\%} compared to WiKiT2M for online scraping and LLM-based prompting, while device coverage shows a marginal decline for the former but increases by \textbf{1.47\%} for the latter. These results indicate that the proposed methods retrieve perceptually similar music.

\begin{table}[h!]
\caption{Subjective evaluation results comparing MTT baseline, online scraping method, and online scraping with LLM-based prompting across three perceptual dimensions (musical similarity, acoustic similarity, and overall similarity).}
\centering
\resizebox{0.45\textwidth}{!}{ 
\begin{tabular}{@{}lcccc@{}}
\hline
\toprule
\multirow{2}{*}{Method} & \multicolumn{4}{c}{Mean Opinion Score (MOS)}  \\ \cmidrule(l){2-5}
& Musical & Acoustic & Overall & Average \\ \hline
MTT \cite{mtt} & 1.79     &  1.80    & 1.93     & 1.84        \\
Online Scraping  &  3.74    &  3.56    & 3.57     &  3.62       \\
\cdashline{1-5}[1pt/2pt]
+ LLM-based Prompting & \textbf{4.44}     &      \textbf{4.07}     & \textbf{3.96} & \textbf{4.16}      \\
\bottomrule
\hline
\end{tabular}
}
\label{tab:mos}    

\end{table}


\begin{table}[h!]
\caption{Online A/B testing results of WikiT2M, online scraping, and online scraping with LLM-based prompting on a commercial streaming platform.}
\centering
\resizebox{0.45\textwidth}{!}{ 
\begin{tabular}{@{}lcc@{}}
\hline
\toprule
Method & $\bigtriangleup$ Playback Count  & $\bigtriangleup$ Device Coverage \\
\midrule
WiKiT2M \cite{WikiMuTe} & - & - \\
Online Scraping & +3.44\% & -0.13\% \\
\cdashline{1-3}[1pt/2pt]
+ LLM-based Prompting & \textbf{+6.20\%} & \textbf{+1.47\%} \\
\bottomrule
\hline
\end{tabular}
}
\label{tab:ab}
\vspace{-0.2cm}
\end{table}

\textbf{Ablation study}, presented in Table \ref{tab:ablation}, evaluates the impact of modifying or replacing textual data on system performance. Experiments are conducted on the CWE and WikiMuTe datasets using captions and aspects. \textbf{a)} It has been observed that songs by the same artist are often prioritized in the retrieval results, which undesirably reduces retrieval diversity. Specifically, out of 5.3k queries in the music-to-music retrieval evaluation, \textbf{42.23\%} retrieve at least one song by the same artist, with an average of 2 songs among the 100 items retrieved. To address this, we experiment with masking titles and names in the captions to encourage the model to focus on the underlying musical content. We observe that the ratio marginally decreases to 41.94\%. It is speculated that the music encoder has internalized strong artist-specific features, such as timbre, which persist even when explicit artist names are removed. \textbf{b)} Replacing online scraped text with GPT-4o-mini generated text (ID.3-4) shows consistent improvements across both coarse and fine levels. This demonstrates that LLM-generated descriptions match human-created scraped text in semantic content and overall quality. More importantly, this indicates the feasibility of constructing a high-quality text-music dataset entirely using LLMs, eliminating the need for laborious online sourcing and mining efforts.

\begin{table}[h!]
\caption{Ablation study of music-to-music retrieval performance w/wo masking strategy and pseudo labeling.}
\centering
\resizebox{0.4\textwidth}{!}{ 
\begin{tabular}{lcccc}
\hline
\toprule
\multirow{2}{*}{ID} & \multirow{2}{*}{Mask} & \multirow{2}{*}{Pseudo-label} & \multicolumn{2}{c}{Music-to-Music} \\ \cmidrule(l){4-5} 
  &   &   &  Coarse  & Fine \\ \hline
1 &  \ding{55} & \ding{55} & 47.09 & 53.35 \\  
2 &  \ding{51} & \ding{55} & 47.12 & 53.27 \\ 
3 &  \ding{55} & \ding{51} & 47.49 & 53.41 \\  
4 &  \ding{51} & \ding{51} & \textbf{47.76} & \textbf{55.00} \\ 
\bottomrule
\hline
\end{tabular}
}
\vspace{-0.4cm}
\label{tab:ablation}
\end{table}

\section{Conclusion}
This paper presented a novel framework for music similarity retrieval that leverages cross-modal learning to overcome traditional limitations in musical relationship modeling. The proposed dual-source approach to text data acquisition, combining online scraping and LLM-based automated generation, effectively addresses the challenge of data scarcity. Extensive experiments demonstrated that our framework significantly outperforms existing benchmarks through objective metrics, subjective evaluations, and real-world A/B testing on a commercial music platform, establishing its practical value for music recommendation systems.





\bibliographystyle{IEEEbib}
\bibliography{icme2025references}

\vspace{12pt}
\end{document}